**Knockdown of human AMPK using the CRISPR-Cas9 genome-editing system**

Adrien Grenier[1,2,3,4,#], Pierre Sujobert[5,6,#], Séverine Olivier[1,2,3,#], Hélène Guermouche[1,2,3,4], Johanna Mondésir[1,2,3,4], Olivier Kosmider[1,2,3,4] Benoit Viollet[1,2,3,§] and Jérôme Tamburini[1,2,3,4,#,§]

[1] Institut National de la Santé et de la Recherche Médicale (INSERM), U1016, Institut Cochin, Département Développement, Reproduction, Cancer, Paris, France.

[2] Centre National de la Recherche Scientifique (CNRS), Unité Mixte de Recherche (UMR) 8104, Paris, France.

[3] Université Paris Descartes, Faculté de Médecine Sorbonne Paris Cité, Paris, France

[4] Equipe Labellisée Ligue Nationale Contre le Cancer (LNCC), Paris, France

[5] Hospices Civils de Lyon, Centre Hospitalier Lyon Sud, Service d'Hématologie Biologique, Pierre-Bénite, France. Université Claude Bernard Lyon-1, Lyon France.

[6] INSERM U1052, CNRS 5286, Université Claude Bernard, Faculté de Médecine Lyon-Sud Charles Mérieux, Université de Lyon, Pierre Bénite, France.

[#] Contributed equally to this work

[§]To whom correspondence should be addressed:

Jérôme Tamburini, INSERM U1016, Institut Cochin, 22 rue Méchain, 75014 Paris, France; phone: +33140516510; fax: +33140516543; email: jerome.tamburini@inserm.fr

Benoit Viollet, INSERM U1016, Institut Cochin, 24 rue du faubourg Saint Jacques, 75014 Paris, France; phone: +33144412401; fax: +33144412421; email: benoit.viollet@inserm.fr

**Running head:** CRISPR-Cas9 for human AMPK

**Summary**

AMP activated protein kinase (AMPK) is a critical energy sensor, regulating signaling networks involved in pathology including metabolic diseases and cancer. This increasingly recognized role of AMPK has prompted tremendous research efforts to develop new pharmacological AMPK activators. To precisely study the role of AMPK, and the specificity and activity of AMPK activators in cellular models, genetic AMPK inactivating tools are required. We report here methods for genetic inactivation of AMPK α1/ α2 catalytic subunits in human cell lines by the CRISPR/Cas9 technology, a recent breakthrough technique for genome editing.

**Key words**

CRISPR/Cas9; AMPK; AMPKα1; PRKAA1; AMPKα2; PRKAA2, AML, Caco-2

# 1. Introduction

## 1.1 CRISPR/Cas9 genome-editing system

The rapid development of the Clustered Regularly Interspaced Short Palindromic Repeats (CRISPR)/ CRISPR associated protein 9 endonuclease (Cas9) technology provides unprecedented genomic research tools. CRISPR are short DNA sequences found in bacteria. Together with the bacterial Cas9 these sequences are involved in the clearance of foreign DNA from bacteria [1]. Shortly after its discovery, the CRISPR/Cas9 system has been engineered for genome editing including in eukaryotic cells. Indeed, CRISPR/Cas9 offers a specific genomic tool based on the complementation of genomic DNA by a 20-mer oligonucleotide sequence followed by a consensus sequence called PAM for protospacer adjacent motif. This guide RNA (sgRNA) allows the recruitment of Cas9 to the targeted genomic DNA sequence leading to DNA double strand breaks (DSB). In case of DSB repair by non homologous end joining (NHEJ), small insertions/deletions within coding sequences may be produced leading to a premature stop codon [2,3]. Alternatively, DSB may be repaired by homologous recombination (HR), which may be exploited for targeted mutagenesis of endogenous genes when using a mutated recombination template [4]. In fact, multiple refinements of the CRISPR/Cas9 technology have been described and the power of these techniques as well as their potential applications is increasingly growing.

**1.2 Why and how disrupting AMPK activity?**

AMP-activated protein kinase (AMPK) is one of the main cellular energy sensor, activated by the binding of AMP or ADP to its γ-regulatory subunit and by the phosphorylation of its α-catalytic subunit [5]. Once activated, AMPK modifies cell metabolism (inhibition of anabolism and activation of catabolism) to restore cellular energy balance. Tremendous research efforts are therefore ongoing to efficiently activate AMPK in the context of metabolic diseases or cancer [5]. Abrogating AMPK activity may help to define the role of AMPK in a given cellular context. Unfortunately, no specific AMPK inhibitors are currently available and genetic knockdown/knockout approaches are recommended to validate the function of AMPK (see Chapter 12). Hence, specific deletion of AMPK genes represents an invaluable tool to assess the specificity of AMPK-targeting small molecules (mostly AMPK activators). While targeting regulatory AMPKβ or γ subunits may be important to disrupt the regulation of AMPK activity and function, AMPKα gene inactivation is sufficient to abrogate AMPK serine/threonine kinase activity [6]. At first glance, we used RNA interference (RNAi) by small hairpin RNA (shRNA) to deplete AMPKα from acute myeloid leukemia (AML) cells as done by other groups in other cellular contexts [7-9]. However, we failed to achieve a complete inhibition of AMPKα, with a residual expression of 10-20% at the protein level [10]. Indeed, Saito and colleagues recently demonstrated the essential pro-survival function of AMPK in AML cells exposed to stress [11], thus explaining the negative selection of cells with efficient AMPK α1 knockdown. Hence, efficient and stable AMPKα knockout in human cell lines represents a challenge. To avoid this limitation, and also to reduce the probability of off-target as observed with shRNA [12], we moved towards a CRISPR/Cas9-based approach to study AMPK function in human cell lines. In this chapter, we describe the use of CRISPR/Cas9 technology to disrupt AMPKα expression in acute myeloid leukemia (AML) MOLM-14 and OCI-AML3 cells, and concomitant AMPKα and AMPKα2 expression in colon carcinoma Caco2 cells that may provide invaluable tools for further studies on AMPK biology and development of specific AMPK activators for the treatment of cancer and metabolic diseases. We attempted to find common DNA sequences across *PRKAA1* and *PRKAA2* genes (encoding the AMPKα1 and AMPKα2 catalytic subunits, respectively) for targeting by a single sgRNA, but no such sequence was available within *PRKAA1* or *PRKAA2* coding or non-coding DNA. While multiplexing sgRNA into a single vector is possible [13], we used a sequential procedure to achieve first AMPKα1 and then AMPKα2 knockdown in single-cell clones. We thus developed *PRKAA1-* and *PRKAA2-*targeting CRISPR/Cas9 systems allowing a convenient sequential selection of AMPKα-depleted single cell clones based on antibiotic resistance, and then isolation of AMPKα2 knockout single-cell clones based on the expression of a fluorescent marker. In particular, we describe procedures to design AMPKα and AMPKα2 targeting sgRNAs, and cloning into lentiviral vector. We also provide methods for infection, clonal selection, screening and validation of single-cell CRISPR clones.

## 2. Material

### 2.1 sgRNA design and cloning

1. Online bioinformatic sgRNA design tool (eg, CRISPR design tool). (See note 3).

2. lentiCRISPR v2 plasmid (Addgene reference #52961. (See notes 4 and 5).

3. pL-CRISPR.EFS.GFP plasmid (Addgene reference #57818).

4. Designed sgRNA oligonucleotides: 100 μM oligonucleotides in H$_2$O.

   <u>- PRKAA1 sgRNA#1:</u> forward 5'-CACCGAGGGCACGCCATACCCTTG-3' and reverse 5'-AAACCAAGGGTATGGCGTGCCCT-3'.

   <u>- PRKAA1 sgRNA#2:</u> forward 5'- CACCGATCCTGAAAGAGTACCATTC -3' and reverse 5'- AAACGAATGGTACTCTTTCAGGAT-3'.

   <u>- PRKAA2 sgRNA#1:</u> forward 5'-CACCGAAGATCGGACACTACGTGCT-3' and reverse 5'-AAACAGCACGTAGTGTCCGATCTTC-3'.

   <u>- PRKAA2 sgRNA#2:</u> forward 5'-CACCGTCAGCCATCTTCGGCGCGCG-3' and reverse 5'-AAACCGCGCGCCGAAGATGGCTGAC-3'.

5. 10X kinase buffer: 500 mM Tris-HCl, pH 7.6, 100 mM MgCl$_2$, 50 mM DTT, 10 mM ATP, 1 mM spermidine.

6. T4 PNK: 10 U/μl T4 polynucleotide kinase.

7. Restriction enzyme: 10 U/μl EspI (BsmBI) and 10X FastDigest green buffer (ThermoFisher scientific).

8. T4 ligase: 1 U/ μl T4 ligase.

9. 10X ligation buffer: 500 mM Tris-HCl, pH 7.5, 100 mM MgCl$_2$, 10 mM ATP, 10 mM DTT, 50% (w/v) polyethylene glycol-8000.

10. Stbl3 *Escherichia coli* competent cells.

11. SOC medium: 20 g/l tryptone, 5 g/l yeast extract, 10 mM NaCl, 2.5 mM KCl, 10 mM MgCl$_2$, 10 mM MgSO$_4$, 20 mM glucose. Dissolve to 1 L deionised water, 20 g of tryptone, 5 g of yeast extract, 2 ml of 5 M NaCl, 2.5 ml of 1 M KCl, 10 ml of 1 M MgCl$_2$, and 10 ml of 1 M MgSO$_4$. Autoclave at 121°C for 15 min and add 20 ml of filter-sterilized 1 M glucose.

12. LB agar plate: lysogeny broth (LB) agar plate with 100 μg/ml ampicillin.

13. Maxiprep plasmid DNA kit.

## 2.2 Production of lentiviral particles

Before starting work with lentivirus, ensure compliance with your Environmental Health and Safety office and government/organization/university.

14. HEK 293T cells.

15. Phosphate buffer saline (PBS): 137 mM NaCl, 2.7 mM KCl, 10 mM Na$_2$HPO$_4$, 1.8 mM KH$_2$PO$_4$, pH 7.4.

16. CaCl$_2$ solution: 1 M CaCl$_2$.

17. Lipofectamine® 2000 transfection reagent.

18. PsPAX2 plasmid (encoding Gag and Pol proteins). (Addgene reference #12260).

19. pMD2.G (encoding VSVG envelope proteins). (Addgene reference #12259).

20. HEPES buffered saline (HBS): 20 mM HEPES, pH 7, 150 mM NaCl.

21. Complete DMEM medium: DMEM, 10% (v/v) foetal bovine serum (FBS), 100 units/ml penicillin and 100 μg/mL streptomycin, 2 mM L-Glutamine.

22. Complete MEM alpha medium: MEM alpha medium, 10% (v/v) foetal bovine serum (FBS), 100 units/ml penicillin and 100 μg/mL streptomycin, 2 mM L-Glutamine.

23. OPTIMEM culture medium.

24. 0.22 μm filter.

25. G418 (Geneticin®) stock solution: 100 mg/ml G418.

## 2.3 Infection and clone selection

26. Human AML cell lines: MOLM-14 or OCI-AML3 human cell lines.

27. Human Caco2 colon carcinoma cell line.

28. Complete MEM alpha medium: Minimum Essential Medium (MEM) alpha medium, 10% (v/v) FBS, 100 units/ml penicillin and 100 μg/mL streptomycin, 2 mM L-Glutamine.

29. Complete EMEM culture medium: Eagle's Minimum Essential Medium (EMEM), 20 % (v/v) FBS, non-essential amino acids, 100 units/ml penicillin and 100 μg/mL streptomycin.

30. Trypsin/ EDTA solution: 0.25% (w/v) Trypsin, 0.53 mM EDTA.

31. 50 μm cell strainer.

32. Cell sorting buffer: PBS (without $Ca^{2+}/Mg^{2+}$), 25 mM HEPES, pH 7, 1% (v/v) FBS, 1 mM EDTA.

33. 12 ml FACS tube

34. Flow cytometer.

## 2.3 Screening

35. GeneScan™ 500ROX

36. PCR Primers:

    - <u>PRKAA1 primers:</u> forward 5'-FAM-ATCACCAGGATCCTTTGGCA-3' and reverse 5'-TGCTTTCCTTACACCTTGGTG-3'.

    - <u>PRKAA2 primers:</u> forward 5'-FAM-GCTGCACTGTGGGTAGGC-3' and reverse 5'-GGGCGTCGGCACCTTC-3'.

37. Capillary electrophoresis.

38. GeneMapper® software v3.7

**2.4 Sequencing**

39. Sequencing primers:

    - PRKAA1 primers: forward 5'-ATCACCAGGATCCTTTGGCA-3' and reverse 5'-TGCTTTCCTTACACCTTGGTG-3'.

    - PRKAA2 primers: forward 5'-GCTGCACTGTGGGTAGGC-3' and reverse 5'-GGGCGTCGGCACCTTC-3'.

40. Cell lysis buffer: 4% Tween 20,100 µg/ml proteinase K in $H_2O$.

41. DNA polymerase: Phire Hot Start II DNA polymerase and 5X reaction buffer (Thermo scientific).

42. Tris acetate EDTA (TAE) buffer: 40 mM Tris-HCl, pH 8.3, 20 mM acetic acid, 1 mM EDTA.

43. Agarose gel: 2% (w/v) agarose in TAE.

**3. Methods**

**3.1. sgRNA design and cloning**

Putative target sites can be identified by simply scanning the region of the particular genomic location that you want to target (but you should check your sgRNA design for potential off-target effects) or by using online bioinformatical tools dedicated to sgRNA design. (See note 3).

Human AMPKα catalytic subunit encompasses two isoforms, AMPKα1 and AMPKα2, encoded by the *PRKAA1* and *PRKAA2* genes located at chromosomes 5p13.1 and 1p32.2, respectively. *PRKAA1* is made of 9 exons (Figure 1A) representing 5088 base pairs (bp), and encodes a 559 aminoacids (AA) protein of 64 kDa. *PRKAA2* coding sequence is 9280 bp-long but with a long 3' untranslated region, and encodes a 552 AA protein of 62 kDa (Figure 1B).

To target *PRKAA1* gene, we designed 2 sgRNAs targeting *PRKAA1* exon 7. To target *PRKAA2* gene, we designed 2 sgRNAs targeting *PRKAA2* exon 1 (Figure 1A and B).

[Figure 1 near here]

CRISPR design

1. Identify sgRNA targeting sequences in the genomic region of interest by running the CRISPR design tool (See notes 6, 7 and 8). The output window shows 23 bp genomic sites of the form 5'-$N_{20}$NGG-3' within your target region. These sites may reside on the + or -strand (Figure 2).

2. Remove NGG sequence to leave 20 nt target sequence.

3. Add G to the 5' end of target sequence if it does not begin with G. (See note 9).

4. For cloning into the plentiCRISPR vector using BbsI restriction enzyme, add GATC to the 5' end of forward oligonucleotide, and add AAAA to the reverse complement oligonucleotide (including any additional G nucleotide). (See note 10).

5. Synthetize forward and reverse oligonucleotides corresponding to the selected sgRNAs. (See notes 11 and 12)

CRISPR cloning

6. Mix 1 µl of 100 µM of forward and reverse oligonucleotides with 1 U of T4 PNK in 10 µl of 1X kinase buffer and incubate for 30 min at 37°C. (See note 13).

7. Incubate at 95°C for 5 min and gradually cool the solution at room temperature.

8. Digest 5µg of plentiCRISPR v2 plasmid backbone with 30 U of Esp3I (BsmBI) in a final volume of 60µl of 1X FastDigest green buffer. Incubate for 25 min at 37°C, and then at 65°C for 15 min. (See note 14)

9. Prepare a 10 µl ligation reaction mix by adding 150 ng of plentiCRISPR digestion product, 1 µl of 1 µM annealed sgRNA oligonucleotides, 1 µl of 1 U/µl T4 ligase, 1µl of 10X T4 ligase buffer. Incubate for 10 to 30 min at room temperature.

10. Transform the ligation product into Stbl3 *Escherichia coli* (See note 15).

11. Incubate ligation product and one aliquot of Stbl3 cells for 10 minutes on ice.

12. Heat-shock at 42°C for 45 seconds and return to ice for 2 min.

13. Add 250 µl of room temperature SOC medium and incubate at 37°C with shaking for 1 h.

14. Spread the transformation mixture onto a prewarmed LB agar plate containing 100 µg/ml kanamycin. Incubate overnight at 37°C.

15. Incubate the plates overnight at 37 °C in a microbiological incubator.

16. After incubation, pick 5 to 10 colonies to identify a correct clone for proper insert identification by Sanger sequencing.

17. After identifying the colonies with the correct sequence, isolate plasmid DNA using a maxiprep kit.

[Figure 2 near here]

### 3.2 Lentivirus production

To produce recombinant lentivirus, HEK293-T packaging cells are transfected with the packaging plasmids pVSVg and psPAX2 encoding lentiviral proteins (Gag, Pol and Env) and the transfer plentiCRISPR/ sg RNA plasmid. We propose here two different methods for lentiviral particles production using HEK293T packaging cells. (See note 16).

Calcium/HBS transfection

18. Plate 1.5x10$^6$ HEK293T cells into two T175 flask in complete DMEM medium containing 1 mg/ml G418 antibiotic (generally 4 days before transfection).

19. The day of transfection, wash cells once with PBS and add 18 ml of DMEM without FBS or G418. Incubate the cells for 6h in a humidified 37°C, 5% CO$_2$ incubator.

20. Prepare transfection solution by adding 1.5 ml of sterile water, 500 µl of 1 M CaCl2 solution and a DNA plasmid mix containing 16 µg pMD2.G, 24 µg PsPAX2 and 3µg of plentiCRISPR/ sgRNA plasmids.

21. During a continuous vortex agitation, add dropwise transfection solution into 2 ml of HBS in a 50 ml tube. Incubated the mixture (final volume of 4 ml) for 15 min at room temperature.

22. Add 2ml of this mix on each T175 flask and incubate overnight in a humidified 37°C, 5% CO$_2$ incubator.

23. Discard culture medium and add 18 ml of complete DMEM culture medium. Incubate in a humidified 37°C, 5% CO$_2$ incubator for 24h.

24. Collect the supernatants using a syringe and filtrate out the cellular debris using a 0.22µm filter. The filtered lentivirus supernatant can be either immediately used to transduce a cell line or stored at -80°C. Avoid freeze-thawing lentivirus supernatant. (See note 17).

Lipofectamine transfection

25. Plate 2x10$^6$ HEK293T cells into two 10 cm diameter culture dishes with DMEM culture medium containing 1 mg/ml G418 antibiotic and incubate overnight in a humidified 37°C, 5% CO$_2$ incubator.

26. Wash once with PBS and replace medium with 4.2 ml of OPTIMEM culture medium.

27. Add 12µl of lipofectamine 2000 transfection reagent to 400µl of FBS-free OPTIMEM medium. Incubate the mixture at room temperature for 15 min.

28. Add 3µg of plentiCRISPR/ sgRNA, 2µg of PsPAX2 vector and 1µg of pMD2.G plasmids to 400µl of FBS-free OPTIMEM medium.

29. Mix diluted plasmid DNA with diluted Lipofectamine mixture. Incubate the mixture at room temperature for 15 min.

30. Add 800 µl of the mixture dropwise to the cells and incubate in a humidified 37°C, 5% $CO_2$ incubator.

31. Replace medium after 3-6 h with fresh DMEM culture medium and incubate in a humidified 37°C, 5% $CO_2$ incubator for 48 h.

32. Collect the supernatants using a syringe and filtrate out the cellular debris using a 0.22µm filter. The filtered lentiviral supernatant can be either immediately used to transduce a cell line or stored at -80°C. Avoid freeze-thawing lentiviral supernatant. (See note 17).

**3.3 Cell infection and clonal isolation**

Determination of MOI

The volume of supernatants used for infection may vary for each cell line and depends on lentiviral titer. Determination of the multiplicity of infection (MOI) for each cell line is recommended and is valid only for a given lentiviral supernatant batch [14]. To ensure that most cells receive only one stably-integrated sgRNA, a MOI of 0.3 should be used. (See note 18).

MOI theoretically is the number of viral particles per cell, a MOI of 1 meaning that 1 particle reached a single cell. However, particles may not lead to viral genome integration in all cases and transduction is not linear but follows a Poisson distribution considering that each event is independent. MOI can be determined by using the following formula:

$m = -ln(1-P)$

where $m$ = MOI and $P$ = probability of infection.

33. Plate 12-well plates at a density of $2 \times 10^5$/ml AML cells in 2 ml of complete MEM alpha medium.

34. In each well, add 200 μl, 100 μl, 50 μl, 25 μl or 0 μl of lentiviral supernatant. An additional no antibiotic selection control well without addition of lentivirus is prepared. Incubate for 24 h in a humidified 37°C, 5% $CO_2$ incubator.

35. 24 h after lentiviral transduction, wash the cells twice with PBS to eliminate residual lentiviral particles, and then incubate with complete MEM alpha medium.

36. 24 h after lentiviral transduction, replace the medium with complete MEM alpha medium containing 1 mg/ml puromycin (or the relevant antibiotic selection). For the no antibiotic selection control well, replace the medium with complete MEM alpha medium without antibiotic.

37. Refresh complete MEM alpha medium with and without antibiotic until the antibiotic selection control well with no virus contains no viable cells.

38. Count the number of cells in each well. For each virus conditions, MOI is estimated. For example, if 37% of transduced cells are recovered after antibiotic selection, MOI = $-\ln(0.37) = 0.994$.

Infection of cell lines

39. For AML human cell lines, seed cells at a density of $2 \times 10^5$ cells/ml in complete MEM alpha medium.

40. For colon carcinoma Caco2 cells, plate $2 \times 10^5$ cells/well in 24-well plate in 0.5 ml of complete EMEM culture medium.

41. Add lentiviral supernatants at a MOI of 0.3 and incubate for 24 h in a humidified 37°C, 5% $CO_2$ incubator.

42. Wash the cells with PBS twice and then incubate with complete MEM alpha medium for AML cells or complete EMEM culture medium for Caco2 cells.

Antibiotic selection and clonal isolation

43. Two to three days after lentiviral transduction, replace the medium with complete MEM alpha medium containing 1 mg/ml puromycin (or relevant antibiotic selection depending on antibiotic resistance marker expressed on the lentivirus vector used) for AML cells or complete EMEM culture medium containing 1 mg/ml puromycin (or relevant antibiotic) for Caco2 cells.

44. After antibiotic selection of transduced cells, select clonal cell lines by isolating single cells into 96-well plates through either cell sorting or other isolation procedures (See notes 19 and 20).

45. For single isolation of AML cells, count cells with an automated cell counter and prepare a cell suspension with $10^6$ cells/ml in cell sorting buffer.

46. For single isolation of Caco2 cells, aspirate culture medium and wash cells with room temperature PBS.

47. Add trypsin/EDTA and incubate for 3 min to detach cells.

48. Resuspend dissociated cells with prewarmed complete EMEM medium and gently pipet up and down several times to generate a single-cell suspension.

49. Centrifuge the cell suspension for 5 min at 800 g at room temperature.

50. Aspirate the supernatant and resuspend the cell pellet in appropriate volume of cell sorting buffer to adjust the cell concentration to $10^6$ cells/ml.

51. Filter cell suspensions through a 50 μm filter into a FACS tube.

52. Set up the cell sorter and install the required collection device.

53. Collect cells individually into 96-well plates containing 100 μl/ well of appropriate culture medium.

54. After isolation of single cells (or single-cell derived clones), culture the cells without selection antibiotic. Allow the bulk cells to expand in a humidified 37°C, 5% $CO_2$ incubator. Establishment of new clonal cell lines will vary depending on the doubling time of the cell line used.

Fluorescence-activated cell sorting (FACS) enrichment and clonal isolation

The following steps are performed for clonal isolation without antibiotic selection when using a lentiCRISPR plasmid expressing a fluorescent marker (eg, pL-CRISPR.EFS.GFP plasmid). This approach can be a useful alternative strategy when working with puromycin (or other antibiotic) resistant cell lines. Indeed, we used FACS to select cells transduced with 'GFP' pL-CRISPR.EFS.GFP/ AMPKα2 in puromycin-resistant AMPKα1 knockout Caco2 cells (previously generated by transduction with 'puromycin' lentiCRISPR/AMPKα1) (See note 21) (Figure 3).

55. Two to three days after lentiviral transduction, confirm successful delivery of sgRNA by visualizing fluorescent marker expression using fluorescence microscopy.

56. Prepare cells for FACS analysis as described in section 3.3 steps 45 to 51 for suspension and adherent cells.

57. Set up the flow cytometer with negative control cells. Adjust the forward scatter (FSC) and side scatter (SSC) to place the population of interest on scale.

58. Run positive control cells and draw an interval gate to define the populations of interest.

59. Once gate have been determined, sort the top ~10% of fluorescent positive cells from the experimental samples.

60. Collect cells individually into 96-well plates containing 100 μl/ well of culture medium.

61. After sorting, examine the plates under a microscope and determine the presence of a

single cell in most of the wells on the plate. Mark off the wells that are empty or that may have been seeded more than a single cell.

62. When cells exceed $10^5$ cells/ml for suspension cells and are 80-90% confluent for adherent cells, prepare replicate plates for screening.
63. Expand the cells for 2–3 weeks. Change medium every 3–5 days as necessary and passage accordingly.

[Figure 3 near here]

### 3.4 Screening and validation

Most mutations induced by sgRNA at target site are small deletions or insertions (± 1 to 20 bp) and knockout cell clones can be simply identified by analysing successful micro-deletions or -insertions by product size analysis. Forward and reverse primers designed to anneal outside of the target locus region are used.

64. Screen a coverage of >20 clones per sgRNA to guarantee that each perturbation will be sufficiently represented in the final screening readout. (See note 22).

<u>Cell lysis</u>

65. Transfer cells into a 96-well U-bottom plate and wash once in PBS buffer through centrifugation at 1200 rpm for 20 seconds at 4°C.
66. Discard culture medium and add 13μl of cell lysis buffer. (See notes 23 and 24).
67. Transfer into PCR tubes, vortex briefly and heat at 56°C for 10 min and then at 95°C for an additional 10 min.

<u>PCR amplification</u>

68. Select a 500-bp region around the targeting site and identify optimal targeting primer sets that amplify 200 to 300 bp around the targeting site. (See note 25).

69. Assemble a 50 µl PCR with 13 µl of cell lysate, 1.5 µl of 10 µM 5'-FAM labeled forward primer, 1.5 µl of 10 µM non-fluorescent reverse primer, 1 µl of 10 mM dNTP, 10 µl of 5X PCR buffer and 1 µl of Phire taq polymerase.

70. Run sample in thermocycler with the following program: 98°C for 1 min and for 35 cycles:, 98°C for 7 sec, X°C for 5 sec (X depend on the annealing temperature of the primers), 72°C for 10-15 sec/ kb, and then 72°C for 1 min. (See note 26).

71. After the reaction is complete, run 1 µl of each amplified target on a 2% (w/v) agarose gel in TAE to verify successful amplification of a single product at the appropriate size.

Fragment analysis

72. To 1µL of fluorescent PCR products, add 18.8 µL of water and 0.2 µL of GeneScan™ 500ROX.

73. Perform migration by capillary electrophoresis using a DNA analyzer.

74. Analyse migration profile using GeneMapper® software v3.7 (Applied Biosystems) (Figure 4C, left panel).

[Figure 4 near here]

Validation of genetic modification by Sanger sequencing

75. Use non-fluorescent primers to generate amplicons as described above (steps 68-70).

76. Sequence the amplicons by using the same primers. Representative chromatograms are provided in Figure 4C (right panel) for correlation with fragment analysis for the same clone.

_Verification of knockout cell clones by immunoblotting analysis_

After validation of sequence perturbation, Western blot analysis is recommended for verifying protein expression of targeted genes (See note 24). Methods for performing Western blot are described in Chapters 24 and 27.

Western blots from CRISPR/AMPKα1-modified MOLM-14- and OCI-AML3-derived clones are shown in Figures 5A-C. Western blots from CRISPR/AMPKα1- and CRISPR/AMPKα2-modified Caco2-derived clones are shown in Figure 3A and B.

[Figure 5 near here]

**4. Notes**

1. Although we provide a comprehensive view of CRISPR/Cas9 use in human AML and colon carcinoma cell lines, the methods described here may be applied to AMPK knockdown in other human cell lines as well as to any gene of interest.

2. Recent breakthrough allows researchers to disrupt the expression of a given gene by multiple means. Among the most popular techniques are RNA interference and CRISPR/Cas9, whose pros and cons are reviewed by other [12]. While both techniques might be used to achieve a partial knockdown of protein expression (for example with inactivated dCas9 fused to transcriptional modulators [15]), only CRISPR/Cas9 can achieve a complete and stable inhibition of protein expression. We believe that most of the technical pitfalls in CRISPR/Cas9 assays could be overcome after a careful adaptation of our current canvas to each cell type and sgRNAs, providing an invaluable tool for AMPK pharmacological research and beyond.

3. Many bioinformatics tools are available online to help in the design of sgRNA (e.g., ZiFit, CRISPR Design Tool). We used the optimized CRISPR Design application software from Dr Feng Zhang's laboratory (http://crispr.mit.edu). This software provides an accurate analysis of off-target sites. Due to the growing use of the CRISPR/Cas9 technology, more and more sgRNA sequences will be published to target a given gene/protein. Thus, it will be possible to pick-up already validated sgRNA sequences into published articles. Some companies also offers ready-to-use plasmids containing a validated sgRNA and also services for lentiviral particle production for a given plasmid that may represents a valuable assets, particularly for teams with limited experience with the CRISPR/Cas9 technology.

4. In previous publication, we have used the lentiCRISPRv1 plasmid [10], formerly referenced under the Addgene reference #49545, but this plasmid is currently not available due to the recent development of lentiCRISPRv2 plasmid (Addgene reference #52961). lentiCRISPRv2 is identical to the original lentiCRISPRv1 but produces nearly 10-fold higher titer virus.

5. We mostly used the LentiCRISPR v1 plasmid, allowing the expression of sgRNA and Cas9 though a unique vector. While we found this system very convenient for our purposes, facilitating the clonal selection steps by antibiotic selection and allowing further transductions with other lentiviral vectors, such as GFP-expressing vectors (pL-CRISPR.EFS.GFP plasmid, Addgene reference #57818), people may want to use different systems dependent on their cellular model. Among many CRISPR/Cas9 tools currently available, we mention doxycycline-inducible vectors, such as for example the pLKO.1-puro U6 sgRNA BfuAI stuffer plasmid (Addgene #50920), using a modified Cas9 inactivated for endonuclease activity (dCas9) and fused to transcription co-repressors such as KRAB, allowing a direct and efficient repression of transcription.

6. Appropriate sgRNA sequences are computationally designed based on known sgRNA targeting rules. The selected genomic target sequence should be unique to the genomic target site and be present immediately upstream the PAM sequence that is necessary for Cas9 recognition. Cas9 nuclease cuts 3-nt upstream of the PAM site. Avoid target sequence with homopolymer stretches (eg, AAAA) and with a GC rich content. The sequence analysis software analyses a target sequence of up to 500 nucleotides and searches for PAM sequences either in the positive or negative strand of the corresponding genomic DNA. Every potential sgRNA is then analysed against a reference genome (eg, human genome hg19), to search for potential off-target sites in the whole genome. The output assigns a score (from 0 to 100) for each guide, according to an algorithm described on the website, and indicates the potential undesirable targets of this sgRNA. As recommended by Zhang's team, sgRNA with a score above 50 are good candidates, provided the off-targets are outside gene regions.

7. sgRNA targeting the first exons of the gene are better candidates, to ensure that the premature stop codon will efficiently disrupt protein function. However, when targeting genes encoding multiple isoforms, we recommend targeting exons shared between all isoforms.

8. It is recommended to check the location of the sgRNA on the genomic sequence of the gene of interest, in order to avoid sgRNA that would be complementary to an exon-exon junction and thus would be inefficient at the genomic level.

9. If the 20 bp sequence does not start with a 'G', a single 'G' nucleotide must be prepended to allow efficient transcription from the RNA Polymerase III U6 promoter.

10. Appropriate sites for restriction enzymes should be added for subcloning purpose. If plentiCRISPR plasmid contain a Esp3I (BsmBI) restriction site, add 'GATC' to the 5' end of the forward sequence and 'AAAA' to the 5' end of reverse complementary sequence. The two oligos should anneal to form the following double strand:

5'-GATCGXXG-3'

3'-CXXCAAAA-5'

11. A control sgRNA that does not match any genomic sequence can be also synthesized. We used the following non-targeted guide as a control: 5'-GTAGGCGCGCCGCTCTCTAC -3' [10].

12. We suggest using two (or more) different sgRNA sequences for each target gene, to be able to conclude that the observed effects are directly due to gene knockout, rather than to off-target activities of the constructs.

13. If phosphorylated oligonucleotides are ordered, the use of T4 PNK is omitted.

14. In order to minimize the risk of self-ligation of the plentiCRISPR plasmid, an additional digestion with 30 U of NsiI in the presence of 3 U of thermosensitive alkaline phosphatase (FastAP, EF0651, Thermo Scientific) is performed.

15. Lentiviral transfer plasmids contain Long-Terminal Repeats (LTRs) and must be transformed into recombination-deficient bacteria. We use Stbl3 *E. coli* (ThermoFisher scientific, C7373-03) although other RecA$^-$ strains may be used as well.

16. The choice between calcium/HBS and lipofectamine$^{TM}$2000 to transduce HEK293 cells is mostly based on the cost/efficiency balance. For large lentiviral productions, we generally use calcium/HBS transductions for cost/efficiency reasons. In case of hard-to-transduce cell line, filtrated and concentrated supernatants should be used in priority. Alternatively, lentiviral production can be delegated to commercial companies. Lentiviral supernatants should be stored at -80°C and freezing/thawing cycles should be avoided by aliquoting the supernatants in small fractions.

17. Lentivirus supernatants can be concentrated by ultracentrifugation to increase transduction efficiency that may be helpful for some cell types. Concentration of lentivirus is achieved through centrifugation in a SW32Ti rotor at 22 000 rpm for 1h30 with the following settings: acceleration 3; deceleration 3. Concentrated supernatants are then stored at -80°C.

18. Researchers should be aware that lentiviral transduction efficacy is widely different across cell line, and that each lentiviral supernatant batch also produces different transduction rates for the same cell line. We recommend the determination of a MOI for each lentiviral supernatant and each cell line used. MOI calculation is based on Poisson's statistics, assuming that cell transduction is a binary process. Indeed, MOI better reflects the true cell transduction rate than quantification of lentiviral particle on cell culture supernatants due to the presence of empty particles. While MOI is ideally evaluated when the vector is directly detected in a given cell, for example by a fluorescent protein (GFP or other), an approximation of efficiently transduced cells is possible using antibiotic selection such as puromycin, approximating that only transduced cells are alive after adjunction of the selection compound.

19. Among multiple critical steps using this technique described here, people should keep in mind that, dependent on cell type, CRISPR/Cas9-induced knockdown may not be apparent in a bulk cellular population, in contrast to other techniques such as RNA interference. Several parameters may explain this fact, among which are variability in lentiviral transduction efficacy, discrepancy in sgRNA efficacy and also cell ploidy that may affect CRISPR/Cas9 efficacy in the presence of multiple copy of a given gene. Moreover, CRISPR/Cas9 may induce multiple gene modifications at the single-cell level, which may be difficult to genetically characterize when working on a bulk population. For these reasons, we recommend to single-cell clone transduced cell lines and to choose the optimal clones based on genetic and immunoblotting characterization.

20. Depending on the response of used cell types to FACS, alternative methods can be performed. Cells can be individually sorted by using limiting dilution. We recommend to plate cells at 0.4 cell/well of 96-well plate. Isolation of clonal cell populations can be also performed by platting the cells at low density to isolate individual single-cell derived clones. Monoclonal colonies can be picked and moved to a flat bottom 96-well plate for expansion. As colony formation could vary among cell types, optimal density of cell population needs to be determined empirically.

21. Beyond genetic and proteomic characterization of AMPK knockdown cells, attention should be given to the evaluation of AMPK activity in these cells. In AML cells, we exploited the fact that only AMPKα1 isoform are expressed, allowing a complete disruption of AMPK activity, as attested for example by the absence of induction of ACC phosphorylation upon stimulation by AMPK activators [10]. In other cell types, co-expression of AMPKα2 even at low levels might represent an issue for achieving a complete inhibition of AMPK. In this situation, strategy of sequential targeting of *PRKAA1* and then *PRKAA2* by different CRISPR/Cas9 vectors might represent an option.

22. In our experience, the probability of inducing DNA alterations by CRISPR/Cas9 is variable depending on the vector, sgRNA and cell line. Therefore, achieving a significant depletion of a target protein on a bulk cellular population by this method is uncommon. In MOLM-14 and OCI-AML3 bulk populations, efficient AMPKα1 knockdown is achieved using AMPK#2 sgRNA but AMPKα1 protein expression is barely affected by AMPK#1 sgRNA in these cells (Figure 3A). We may extrapolate the probability of efficient knockdown, as in OCI-AML3 cells, 2/12 (16%) clones were AMPKα1-depleted with AMPK#1 guide and 7/12 (58%) with AMPK#2 guide (Figure 3B). Similar, we found 1/5 (20%) modified clones with AMPK#1 and 5/7 (71%) with AMPK#2 in the MOLM-14 cell line (Figure 3C). We thus suggest performing a single-cell selection of cellular clones that will be further genetically characterized.

23. We have developed an easy-to-use assay to amplify a region avoiding DNA extraction, and using a limited number of cells. This tool allows searching for genomic DNA modifications in a large number of single cell-derived clones simultaneously. However, any method for genomic DNA isolation may be utilized.

24. To screen for multiple clones in parallel, we set up an assay based on direct PCR amplification of our target genomic sequence among cells in liquid cultures. By direct cell lysis into the PCR tube, we avoid time-consuming steps of DNA extraction and purification. While we focused on capillary electrophoresis DNA fragment analysis, as we found this technique robust and cost-efficient for our purposes, other techniques may be applied to characterize CRISPR/Cas9-induced genomic modifications including next-generation sequencing [16]. Although time-consuming, it is possible to search for protein modification across the clones using immunoblotting. Alternatively, we may imagine to directly assess protein expression in single cell clones using reverse phase protein microarrays (RPPA) that allows to test in parallel up to 100 conditions in a single slide [17]. However in this latter case, we still advise to characterize the selected clones at the genetic level as well.

25. Amplification of fragment smaller than 400 bp is recommended to ensure good resolution of the fluorescence-labelled DNA size analysis. Note that the size of amplicon analysis could vary by ~1 bp.

26. To estimate the appropriate annealing temperature for primer pairs when using the Phire DNA polymerase, we use the Tm calculator application available at https://www.thermofisher.com/us/en/home/brands/thermo-scientific/molecular-

biology/molecular-biology-learning-center/molecular-biology-resource-

library/thermo-scientific-web-tools/tm-calculator.html.

**Acknowledgements**

Work from the authors was performed within the Département Hospitalo-Universitaire (DHU) AUToimmune and HORmonal diseaseS (AUTHORS) and was supported by grants from INSERM, CNRS, Université Paris Descartes, and Société Francophone du Diabète (SFD). J.M. was supported by a fellowship from AP-HP. A.G. holds a doctoral fellowship from CARPEM. S.O. received a doctoral fellowship from the Région Ile-de-France (CORDDIM).

**Figure captions**

**Figure 1. Graphical representation of *PRKAA1* and *PRKAA2* genomic DNA sequences.** *PRKAA1* and *PRKAA2* encompass 9 and 10 exons, respectively, Exons are figured as vertical black rectangles. *PRKAA1 e*xon 7 is highlighted in red. Intronic regions are depicted by connecting roof-shaped lines. 3' untranslated regions (3'UTR) are indicated. The scale is indicated, 1Kb: $10^3$ bases. (A) AMPKα1 sgRNA#1 (green) and sgRNA#2 (red); forward (blue) and reverse (pink) sequencing primers; and predicted amplicon (underlined) are represented. (B) AMPKα2 sgRNA 1 (green) and 2 (red); forward (blue) and reverse (pink) sequencing primers; and predicted amplicon (underlined) are represented.

**Figure 2. Screenshots from online CRISPR design tool.** (A) Graphical representation of location for proposed AMPKα1 sgRNA sequence within *PRKAA1* gene. (B) Specificity scores for proposed AMPKα1 sgRNA sequence. Each guide has a specificity score from 0 to

100. (C) Details about potential off-targets. CRISPR Design application software is available at http://crispr.mit.edu.

**Figure 3. Experimental workflow for genome engineering of colon carcinoma Caco2 cells.** We used a sequential procedure to achieve both AMPKα1 and AMPKα2 knockout in single-cell Caco2 clones. (A) PRKAA1-targeting sgRNAs are cloned into the pLenti-CRISPR plasmid, allowing a selection based on puromycin resistance. PRKAA2-targeting sgRNAs are cloned into the pL-CRISPR.EFS.GFP plasmid allowing a selection based on GFP detection. Single-cell sgRNA AMPKα1 Caco2 clones were selected based on puromycin resistance (left part) and then single-cell sgRNA AMPKα2 Caco2 clones on a AMPKα1 knockout background are isolated based on GFP fluorescence (right part). (B) Genome editing efficiency of AMPKα2 sgRNA in AMPKα1-knockout Caco-2 cells is assessed by capillary electrophoresis fragment analysis. (C) Western blots analysis of AMPKα protein expression in single-cell Caco2 clones isolated after sequential transduction with lentivirus expressing 'puromycin' CRISPR/AMPKα1 and 'GFP' CRISPR/AMPKα2. Clone numbers highlighted in red and in black indicate successful AMPKα2 knockout and knockout failure, respectively. Western blots were performed using anti-panAMPKα and anti-β-actin antibodies.

**Figure 4. Fragment and sequencing analysis of single-cell sgRNA AMPKα1 AML clones.** (A) Limiting dilutions experiments in MOLM-14 cells to assess the sensitivity of $H_2O$/tween lysis and AMPKα1 targeted locus amplification. (B) Agarose gel migration of PCR products (using the sequencing primers described in Figure 1A) from single-cell MOLM-14 and OCI-AML3 clones isolated after transduction with lentivirus expressing CRISPR/AMPKα1#1 and #2. (C) Capillary electrophoresis fragment analysis using FAM-labelled primers in OCI-AML3 CTR, OCI-AML3 AMPKα1#2_5 and MOLM-14 AMPKα1#2_15 clones (left panel)

and matching Sanger sequencing for OCI-AML3 AMPKα1#2_5 and MOLM-14 AMPKα1#2_15 clones (right panel).

**Figure 5. AMPKα1 protein expression in sgRNA AMPKα1 AML cell lines.** (A) Western blot analysis of AMPKα1 protein expression in MOLM-14 and OCI-AML3 cell lines bulk populations transduced with lentivirus expressing CRISPR/AMPKα1#1 and #2. (B) Western blots analysis of AMPKα1 protein expression in single-cell OCI-AML3 clones after transduction with lentivirus expressing CRISPR/AMPKα1#1 and #2. (C) Western blots analysis of AMPKα1 protein expression in single-cell MOLM-14 clones isolated after transduction with lentivirus expressing CRISPR/AMPKα1#1 and #2. Clone numbers highlighted in red and in black indicate successful AMPKα1 knockout and knockout failure, respectively. Western blots were performea using anti-AMPKα1 and anti-α-actin antibodies.